**Title:**

Breakdown and polarization contrasts in ferroelectric devices observed by *operando* laser-based photoemission electron microscopy with the AC/DC electrical characterization system

**Authors:**

Hirokazu Fujiwara[1,2*], Yuki Itoya[3], Masaharu Kobayashi[4], Cédric Bareille[5], and Toshiyuki Taniuchi[1,2]

**Affiliations:**

[1]*Department of Advanced Materials Science, Graduate School of Frontier Sciences, The University of Tokyo, Chiba 277-8561, Japan*

[2]*Material Innovation Research Center (MIRC), The University of Tokyo, Chiba 277-8561, Japan*

[3]*Institute of Industrial Science, The University of Tokyo, Tokyo 153-8505, Japan*

[4]*System Design Research Center (d.lab), School of Engineering, The University of Tokyo, Tokyo 153-8505, Japan*

[5]*Institute for Solid State Physics, The University of Tokyo, Chiba 277-8581, Japan*

---

* E-mail: hfujiwara@issp.u-tokyo.ac.jp

**Abstract:**

We have developed an *operando* laser-based photoemission electron microscope (laser-PEEM) with a ferroelectric characterization system. A Sawyer-Tower circuit was implemented to measure the polarization–voltage ($P$–$V$) characteristics of ferroelectric devices. Using this system, we successfully obtained the well-defined $P$–$V$ hysteresis loops for a ferroelectric capacitor incorporating $Hf_{0.5}Zr_{0.5}O_2$ (HZO), reproducing the typical field-cycling characteristics of HZO capacitors. After dielectric breakdown caused by field-cycling stress, we visualized a conduction filament through the top electrode without any destructive processing. Additionally, we successfully observed polarization contrast through the top electrode of an oxide semiconductor ($InZnO_x$). These results indicate that our *operando* laser-PEEM system is a powerful tool for visualizing conduction filaments after dielectric breakdown, the ferroelectric polarization contrasts, and electronic state distribution of materials implemented in ferroelectric devices, including ferroelectric field-effect transistors and ferroelectric tunnel junctions.

**I. Introduction**

Hafnium dioxide (HfO$_2$)-based ferroelectric materials have attracted attention owing to their low switching current, complementary metal-oxide-semiconductor (CMOS)-compatible process, and scalability potential [1–3]. These features enable ferroelectric memories such as ferroelectric random-access memories (FeRAMs)[4,5], ferroelectric field effect transistors (FeFETs)[6,7], and ferroelectric tunnel junctions (FTJs)[8,9] embedded in CMOS large-scale integration (LSI) circuits.

HfO$_2$-based ferroelectric devices face significant reliability challenges. The remanent polarization ($P_r$), which determines the memory window in ferroelectric memory devices, initially increases (wake-up) and subsequently decreases (fatigue) with repeated external field-induced polarization switching [10–12]. The field-cycling characteristics have been attributed to changes in oxygen defect distribution, increase in oxygen vacancies, and/or phase transformations [10,13–15]. Although X-ray photoelectron spectroscopy[16] and transmission electron microscopy (TEM)[15] have provided evidence supporting these mechanisms, their application requires thinning (or removal) of the top electrode (TE) or specimen preparation for cross-section observation. These destructive processes introduce uncertainty in evaluating the defect state, making it difficult to establish a clear correlation between $P_r$ changes and defect state distribution. In recent years, advances in hard X-ray photoemission spectroscopy (HAXPES) have enabled the investigation of chemical states without destructive processes [17–20]. However, microscopic observation has not yet been achieved, and challenges remain in elucidating nanoscale inhomogeneities within devices, which can have a significant impact on their performance.

*Operando* laser-based photoelectron emission microscopy (*operando* laser-PEEM) is a promising technique as a non-destructive imaging method of electronic state distribution in a material buried under an electrode [21–23]. Using an excitation light with energy comparable to the work function of materials, the inelastic mean free path (IMFP) of photoelectrons is theoretically estimated to range from a few nanometers to several tens of nanometers, although it strongly depends on the material [24–26]. In *operando* laser-PEEM studies on resistive random-access memory (ReRAM) using a continuous wave laser with a wavelength of 266 nm as the excitation light, the IMFP in Pt is experimentally estimated to be 16 nm, and the filament responsible for the low resistance state can be visualized through TE with 20 nm thickness [21].

In this paper, we report on the laser-PEEM system with an AC/ DC Electrical Characterization System (ADECS) for evaluating the polarization of HfO$_2$-based ferroelectrics implemented in capacitor devices. In ADECS, a Sawyer-Tower (ST) circuit is implemented to measure polarization–voltage ($P$–$V$) characteristics. $P$–$V$ hysteresis loops and field-cycling characteristics consistent with those reported previously were observed for HfO$_2$-based ferroelectric capacitors by ACECS. Furthermore, we succeeded in visualizing the breakdown spot after the ferroelectric capacitor was broken down. In addition, we successfully observed polarization contrast through TE of an oxide semiconductor. These results demonstrate that our *operando* laser-PEEM system is highly useful for non-destructive analysis of functional materials buried under TE.

## II. Methods

### II.A Sample fabrication

The sample has a crossbar-type structure with 10-nm thick $Hf_{0.5}Zr_{0.5}O_2$ (HZO) as the ferroelectric layer. We fabricated the capacitors on an n-type Si substrate. A 200 nm $SiO_2$ insulating layer was deposited on the Si substrate by radio frequency (RF) reactive sputtering. Metal-ferroelectric-metal capacitors were formed in the stack of TiN/HZO/TiN. The bottom electrode (BE) of the capacitor was formed by depositing TiN by RF sputtering and patterning. The 10-nm HZO was deposited by atomic layer deposition (ALD) as a ferroelectric layer at 250 °C, in which $Hf[N(CH_3)_2]_4$ and $Zr[N(CH_3)_2]_4$ precursors were heated to 75 °C and $H_2O$ was used as the oxidant. The Zr/Hf ratio is 1/1. After ALD deposition, TE was formed by depositing 30-nm TiN by RF sputtering and patterning. For the samples used in polarization contrast observations, $InZnO_x$ (IZO) was deposited by ALD in place of TiN. Rapid thermal annealing at 500 °C for 30 s under $N_2$ atmosphere was performed in order to stabilize the polar orthorhombic phase.

### II.B Laser-PEEM system with ADECS

Our PEEM system uses an aberration-corrected spectroscopic photoemission and low-energy electron microscope (SPELEEM) (Elmitec GmbH) with an energy analyzer as described elsewhere [27]. PEEM imaging was performed using a commercial continuous-wave (CW) laser with a wavelength of 266 nm (4.66 eV photon energy, OXIDE Frequad-M), hereafter referred to as the 266-nm laser. For spectromicroscopy, we also used a CW laser with a wavelength of 213 nm (5.82 eV photon energy, OXIDE Frequad-W), hereafter referred to as 213-nm laser. A sample holder was modified to enable field application to the sample.

To evaluate spontaneous polarization, we used a Sawyer-Tower (ST) circuit, in which a standard capacitor ($C_{st}$) which is connected to the device under test (DUV)[28]. A simplified circuit diagram is shown in Fig. 1(a). Applying drive voltage $V_d$ to the ST circuit, the charges $Q_1$ and $Q_2$ are accumulated in DUT and the standard capacitor, respectively. Since the region between the BE of DUT and TE of the standard capacitor floats electrically, $Q_1$ is equal to $Q_2$; we can define $Q$ as $Q \equiv Q_1 = Q_2$. $Q$ can be obtained from $Q = C_{st}V_{st}$ where $V_{st}$ which is measured across $C_{st}$ by the oscilloscope. The polarization $P$ can be easily obtained by $P = Q/A$, where $A$ is area of DUT. In addition, $C_{st}$ should be chosen much greater than $C_{DUT}$ so that voltage drop across $C_{st}$ is much less than that across $C_{DUT}$. Therefore, the voltage $V_d$ is almost equal to voltage across $C_{DUT}$. The ferroelectric polarization hysteresis loop can be drawn by plotting $P$ versus $V_d$.

For the actual system, however, we need to consider the finite resistance $R_{DUT}$ representative of leakage currents through DUT, the resistor $R_{OSC}$ symbolizing the input resistance of the oscilloscope channel measuring $V_{st}(t)$, and parasitic capacitances $C_{para}$ and $C'_{para}$ as shown in Fig. 1(b) [29,30]. The $C_{para}$ and $C'_{para}$ must be estimated because $C_{st}$ should be sufficiently larger than $C_{para}$ because the voltage $V_d$ needs to be almost equal to the voltage across $C_{DUT}$. When the parasitic capacitance $C'_{para}$ is connected in parallel, the formula to calculate $Q$ is modified as follows: $Q = C_{st}V_{st} \rightarrow Q' = (C_{st} + C'_{para})V_{st}$. $C'_{para}$ can be estimated by fitting $C_{st}$ dependence of $V_{st}^{-1}$ to $C_{st} = Q/V_{st} - C'_{para}$. The ratio $Q/Q'$ indicates the magnitude of underestimation of $Q$ due to the finite $C'_{para}$.

The circuit diagram of laser-PEEM system with ADECS is shown in Fig. 2. For measurements of $P$–$V$ hysteresis loops, a function generator (Keysight 33521B), which can output arbitrary waveform, and an oscilloscope (Keysight DSOX1204A) are used. The TE and BE of the sample capacitor are connected to the electrical system through the copper wires bonded by silver paste. The samples were insulated from the sample holder by Kapton tapes in order to be connected to the standard capacitor directly without connecting the –20-kV potential. Depending on the combination of the states of SW1 and SW2, we can select three modes: (i) Spectroscopic mode using the start voltage (STV), (ii) DC-based measurement mode using the source measure unit, and (iii) AC-based measurement mode using the ST circuit. During PEEM observation, both TE and BE of the sample were connected to the STV. By SW3, we can switch between $C_{st}$ and $C_{BS}$ which is used when we apply bipolar stress to DUT by the function generator in the AC-based measurement mode. We used a $C_{BS}$ with a capacitance significantly larger than the optimized $C_{st}$, ensuring that the voltage drop across the DUT was nearly equal to the output of the function generator.

We first optimized $C_{st}$ for the sample whose designed area was 100 μm × 100 μm. Figures 3(a) and 3(b) show the $C_{st}$ dependence of the ferroelectric characteristics of DUT. The $P$–$V$ curve for $C_{st}$ = 50 pF did not show hysteresis, while that for $C_{st}$ = 2 nF showed a small remanent polarization $P_r$. In contrast, the $P$–$V$ curves for $C_{st}$ = 20 nF and $C_{st}$ = 100 nF showed the clear hysteresis and reasonable $2P_r$ larger than 20 μC/cm$^2$. Therefore, we determined the optimal $C_{st}$ to be at least larger than 20 nF. The lack of observable polarization for $C_{st}$ = 50 pF is due to insufficient voltage applied to the DUT, preventing it from exceeding the coercive electric field. The capacitance of DUT can be estimated to be 266 pF if we assume the linear relative permittivity $\varepsilon$ of 30 for HZO [31]. Ignoring parasitic capacitances, we expect that only 0.63 V of 4 V of the drive voltage is applied to DUT, which is too small to reverse the ferroelectric polarization of 10-nm thick HZO.

To more accurately determine the spontaneous polarization value, the parasitic capacitances must be estimated. $C'_{para}$ can be estimated by fitting the $C_{st}$ vs. $V_{st}^{-1}$ data as shown in Fig. 3(c). From the fitting result, $C'_{para}$ was estimated to be 3.2 nF. $C_{para}$ can be also estimated from the DUT-area dependence of the para-electric capacitance $C_{DUT,p}$ which was obtained from the slopes of the saturated polarization region of $P$–$V$ hysteresis loops. From Fig. 4(a), the slope of the $P$–$V$ hysteresis loops increases with decreasing the area of DUT. As shown in Fig. 4(b), by extrapolating the fitting line of $C_{DUT,p}$ to the point where the DUT area is zero, $C_{para}$ was estimated to be 0.5 nF.

By substituting the determined $C'_{para}$ into the equation $Q/Q' = C_{st}/(C_{st} + C'_{para})$, magnitude of underestimation of $P_r$ can be calculated. In the case of $C_{st}$ = 2 nF, $Q/Q'$ can be estimated to be 0.4, which indicates that the $2P_r$ value shown in Fig. 3(b) is underestimated by 40%. In the case of $C_{st}$ larger than 100 nF, the factor becomes less than 0.03 which corresponds to 0.6 μC/cm² for a polarization of 20 μC/cm². This can be negligible in many cases because changes of $P_r$ induced by field-cycling stress occur on a scale of 1–10 μC/cm². However, if too large $C_{st}$ is used, the signal-to-noise ratio of $V_{st}$ becomes small. Therefore, within the range of acceptable underestimates, an optimal $C_{st}$ should be used.

The frequency of the drive voltage $V_d$ should be also optimized to obtain intrinsic P–V hysteresis loops of DUT. For the measurements of field-cycling characteristics, the frequency of $V_d$ used to measure P–V hysteresis loops should be as high as possible, because the measurements themself can affect the field-cycling characteristics as field stresses, especially in the low cycle number region. Nevertheless, if the frequency is too high relative to the time constant $\tau$ of the CR circuit, the intrinsic P–V hysteresis loop cannot be obtained. Figure 5 shows the frequency dependence of the P–V hysteresis loops of DUT with an area of 100 μm × 100 μm taken for $C_{st}$ of 20 nF. The hysteresis loops taken with 500 Hz and 1 kHz are almost overlapped, while the hysteresis loop taken with 10 kHz shows the higher coercive voltage $V_c$ and distortions near ±3 V which are due to delay of $V_{st}$ relative to $V_d$, namely, due to the phase shift of $V_{st}$ caused by using the high frequency. Therefore, from Fig. 5, the frequency of 1 kHz can be regarded as optimal for our setup.

## III. Results and Discussion

### III.A. Observation of fatigue and breakdown process in ferroelectric capacitors

Using the *operando* laser-PEEM system with $C_{st} = 20$ nF, we observed the fatigue and breakdown process of the ferroelectric capacitor as shown in Fig. 6. The designed sample area is 40 μm × 30 μm. The contribution of leakage current of DUT and parasitic capacitance $C_{para}$ in the observed $V_{st}$ can be removed by the positive-up-negative-down (PUND) method [32,33]. As shown in Figs. 6(a) and 6(b), after 1 cycle of the bipolar field stress was applied, $2P_r$ was obtained to be 47 μC/cm² which is a typical value of $HfO_2$-based ferroelectric capacitors [10–12,34]. The P–V hysteresis loop of the pristine capacitor exhibits a hump-like feature near –1 V. This feature disappears after the application of cycling stress, which is a typical manifestation of the wake-up effect [35]. This indicates that our ADECS works appropriately for the P–V hysteresis curve measurements.

Over the range of 1×10¹ to 1×10⁶ cycles, $2P_r$ gradually decreased from 47 μC/cm² to 39 μC/cm². This behavior is consistent with typical fatigue degradation, indicating that the field-cycling stress is certainly applied to DUT even in conditions where the high voltage is applied between the sample and the objective lens.

At $5 \times 10^7$ cycles, leakage current started to increase, which is a sign of hard dielectric breakdown. At this stage, the P–V hysteresis loop has a distorted shape in the negative electric field side because of the critical leakage current that cannot be removed by the PUND method. At $7 \times 10^7$ cycles, a hard dielectric breakdown occurred. The number of field-cycling stress cycles leading to dielectric breakdown is consistent with the typical endurance of a 10-nm-thick HZO capacitor.

Next, we discuss changes in the PEEM images associated with characteristic modulations of HZO. Figures 6(c)–(f) show the PEEM images taken after $1 \times 10^2$ cycles, $1 \times 10^6$ cycles, $5 \times 10^7$ cycles, and $7 \times 10^7$ cycles, respectively. The raw PEEM images show no significant change in the capacitor region where TE and BE overlap. In order to enhance the slight contrast change, difference PEEM images are shown in Figs. 6(g)–(i). For the difference analysis, intensity of the PEEM images was normalized by the intensity averaged within the area enclosed by solid black squares.

In the region on the capacitor, several-percent difference signals over the entire capacitor area are obtained as light-red and light-blue contrasts. The insets in Figs. 6(g)–(i) show enlarged images of the area enclosed by the dotted black squares in the main images. This field of view includes the distinctive structure of the capacitor, enabling comparison at identical positions on the device. The orange dashed line in the insets shows a certain light-red differential signal region in Fig. 6(g). Within the orange dashed line, the light-red contrast is always formed in each differential PEEM image, indicating that this contrast is independent of the number of the stress cycles after $1 \times 10^6$ cycles. This might suggest that the increase in intensity is derived from fatigue until $1 \times 10^6$ cycles.

Focusing on the green dashed line in the inset, while it is found that only a light-red region exists in Fig. 6(g), light-blue regions coexist in Figs. 6(h) and 6(i). Since the contrast change continued to change even in the $10^7$-cycle range, this contrast change is due to the increase in leakage current, which became critical in the $10^7$-cycle range.

Outside of the capacitor region, the horizontal stripe pattern was observed with large different signals. From the raw PEEM images, the edge pattern of TE is observed unclearly due to charge-up effects. Since in the sample region without TE and BE, 200-nm-thick $SiO_2$ insulates HZO from the Si substrate, the charges in HZO induced by photoemission is not completely compensated. Therefore, the horizontal stripe pattern indicates that the magnitude of the charge-up effect was somehow changed between $1 \times 10^2$ cycles and $1 \times 10^6$ cycles by the stress application.

It is also found that the photoelectron intensity monotonically decreases in the area where HZO is bared on both side of the capacitor region. This area should not be changed by the field-cycling stress because the lateral width between the TEs of DUT and the reference capacitor is 5 μm, and the electric field applied to the bare-HZO region is less than 1/100 of the electric field applied to the capacitor region. The intensity suppression is possibly due to aging originating from laser irradiation for the PEEM image acquisition.

After the hard dielectric breakdown, we clearly observed a region of decreased photoelectron intensity, indicated by a gray arrow in Fig. 6(i). Since the spot appeared just after the leakage current jump was observed, the spot is responsible for the main leakage path formed by the dielectric breakdown. This demonstrates that laser-PEEM is a powerful tool for visualizing the broken spots without any destructive treatment such as removing TE.

In order to investigate the electronic states of the leakage path, energy-filtered PEEM measurements were performed as shown in Fig. 7. From the energy-filtered PEEM images shown in Figs. 7(a) and 7(b), the spot was observed as a dark spot in the low-energy region, while it was observed as a bright spot in the high-energy region. The energy distribution curves (EDCs) extracted at Pos.1 and Pos.2 shown in Fig. 7(a) are shown in Fig. 7(c). The cutoff energy corresponding to $E$–$E_F$ ~ –2.0 eV for the EDCs the 213-nm laser does not show position dependence, indicating that the work function of the TiN surface was not changed by the breakdown event.

However, it was found that the steepness was quite broadened at Pos.1. Rather than reflecting a difference in the intrinsic electronic state of HZO, it is more reasonable to regard this intensity decrease as an effect specific to low-energy electrons with kinetic energy below 0.5 eV. This is because the intensity decrease observed in the EDCs measured with the 266-nm laser is not observed in those measured with the 213-nm laser, and the energy region of the decrease in intensity locates in the lower energy side corresponding to kinetic energy below 0.5 eV. It is expected that the area around the breakdown spot has an internal electric field due to the distribution of chemical potential caused by the drastic change in the electronic state of HZO [22]. The low kinetic energy photoelectrons are strongly affected by this internal electric field, resulting in a defection of their trajectory, and the number of photoelectrons emitted into the vacuum decreased, which might reduce the photoelectron intensity.

In the higher energy region, in contrast, the intensity at Pos.1 is higher than that at Pos.2. In particular, the intensity in the vicinity of the Fermi level ($E_F$), which determines the electrical conductivity, has clearly increased, and the Fermi edge was observed in the difference spectra by both 213-nm and 266-nm CW lasers. Although element-specific analysis is not possible with these results, the findings demonstrate that laser-PEEM can clearly probe the electronic states near $E_F$.

The differential spectral intensity obtained using the 213-nm laser is approximately one-tenth of that obtained with the 266-nm laser. One possible explanation is intensity modulation arising from the photoexcitation process. In photoemission using lower photon energies, it is known that the accessible final states for photoelectrons are limited, leading to intensity modulation—a phenomenon referred to as the final-state effect. However, first-principles calculations show that the conduction band, composed of *p*- and *d*-orbitals, extends over an energy range of approximately 5 eV, and there are abundant final states available for the excitation of non-dispersive in-gap states [36]. Therefore, the final-state effect is not expected to be particularly pronounced in this case.

In addition, the photoexcitation cross-section is known to exhibit photon-energy dependence. According to Ref. 37, the cross-sections of Hf 5d and Zr 4d states increase as the photon energy rises from 10 eV to 15 eV. If this trend continues below 10 eV, it cannot account for the wavelength dependence observed in Fig. 7(c).

The IMFP whose initial states located near $E_F$ differs between 213-nm and 266-nm lasers: IMFP is longer for the 266-nm laser according to the universal curve,[24] that is, a 266-nm laser is more bulk-sensitive than a 213-nm laser. In Fig. 7(c), the amount of the differential signal near $E_F$ is larger for EDCs observed with the 266-nm laser than for EDCs observed with the 213-nm laser. This indicates that the signal near $E_F$ is attributed to the electronic states in the deeper region from the sample surface, supporting the concept that the electronic state of the material buried in TE is observed.

The contribution of photoelectrons originating from regions deeper than the top electrode is quantitatively estimated as follows. The depth-dependent photoelectron intensity can be expressed as $I(z) = A\exp(-z/\lambda)$, where A, $\lambda$, and z indicate a normalization constant, IMFP of the photoelectrons, and the depth from the sample surface, respectively. The fractional contribution $I_d$ from regions deeper than a certain depth $d$ to the total photoelectron intensity is given by the following expression:

$$I_d = \frac{\int_d^\infty exp(-z/\lambda)dz}{\int_0^\infty exp(-z/\lambda)dz} \tag{1}$$

Referring to the universal curve [24], the IMFP is estimated to be $\lambda \sim 6$ nm for the 266-nm laser and $\lambda \sim 4$ nm for the 213-nm laser. The thickness of TE is 30 nm, i.e., $d = 30$ nm, resulting in estimated values of $I_{30nm} \sim 0.6\%$ for the 266-nm laser and $\sim 0.05\%$ for the 213-nm laser.

Although these absolute values depend on the density of defect states and do not fully match the differential intensities shown in Fig. 7(c), the ratio between them is approximately 0.08. This is in rough agreement with the ratio of the differential signals for the 266-nm and 213-nm lasers shown in Fig. 7(c). These results indicate that the primary factor underlying the wavelength-dependence variation in differential intensity is the modulation of the IMFP.

**III.B. Observation of polarization contrast**

We observed polarization contrast of IZO/HZO/TiN capacitors whose cross-section structure is shown in Fig. 8(a). The measurement sequence is shown in Fig. 8(b): after acquiring a $PV$ curve, the polarization reversal and PEEM observation were repeated. The polarization contrast was extracted from the differential PEEM image obtained in these measurements. The same measurement was repeated to validate reproducibility.

Figure 8(c) shows a $P-V$ hysteresis curve of the IZO/HZO/TiN capacitor. From this curve, it is found that the spontaneous polarization of HZO can be reversed by an external voltage of $\pm 3$ V.

From Fig. 8(d), when the PEEM image acquired for the $P_r^+$ state is subtracted from the PEEM image acquired for the $P_r^-$ state, a red contrast is formed in the capacitor region, while when the PEEM image acquired in the $P_r^-$ state is subtracted from the PEEM image acquired in the $P_r^+$ state, a blue contrast is formed. The contrast trend was observed in all the difference images obtained in these measurements, validating the reproducibility. This result indicates that the photoelectron intensity is higher in the $P_r^+$ state than in the $P_r^-$ state.

There are two possible origins explaining the polarization contrast: one is carrier density modulation in IZO. Figures 9(a) and 9(b) show a schematic of the charge distribution in the capacitor. For the $P_r^+$ case, IZO, an n-type semiconductor, is in an accumulation state and the electron density increases, resulting that the photoelectron intensity should increase. This is, however, inconsistent with our experimental results.

Another origin is a modulation of IMFP of photoelectrons. The inner electric field in HZO is not completely screened by IZO because IZO is a degenerate semiconductor and its carrier density is smaller than that of typical metals such as Cu [38]. The finite inner electric field is also confirmed by *operando* HAXPES [17,18]. In the $P_r^+$ state, photoelectrons excited from BE are decelerated by the internal electric field in HZO, resulting in a longer IMFP and increased intensity from BE. In contrast, in the $P_r^-$ state, the photoelectrons are accelerated, and IMFP is shortened, resulting that the photoelectron intensity from BE decreases. This model is consistent with our experimental results.

A more quantitative discussion is provided below to elucidate the polarization contrast mechanism. In the scenario of carrier density modulation by polarization charge, the surface charge density required to compensate a remanent polarization of $P_r \sim 20$ μC/cm$^2$ is estimated to be $P_r/e \sim 1 \times 10^{14}$ /cm$^2$, where $e$ is the elementary charge. This corresponds to only 0.02% of the sheet carrier density in the degenerate semiconductor IZO ($5 \times 10^{17}$ /cm$^2$) [39], and is therefore insufficient to account for the observed ∼1% signal contrast in the differential PEEM images.

Regarding the modulation of IMFP of photoelectrons by the internal electric field, no experimental data for high-carrier-density IZO are currently available. Here, we adopt a reference value of 16 nm, as experimentally estimated for Pt in Ref. 21. A previous report has shown that polarization switching in HZO can induce a modulation in the internal electric field of approximately 1 MV/cm, which corresponds to a 1 eV shift in the photoelectron energy for a film thickness of 10 nm [17]. Assuming that this field modulates the IMFP slightly from 16.0 nm to 15.5 nm, the contribution to the total PEEM signal intensity from depths beyond 20 nm is expected to decrease from 27.7% to 26.6%, as shown in Fig. 9(c). This results in an estimated ∼1% differential signal, which is in reasonable agreement with the experimental contrast observed in Fig. 8.

This quantitative analysis thus supports the interpretation that the polarization contrast primarily originates from modulation of the photoelectron IMFP due to acceleration or deceleration by the internal electric field.

However, the above discussion involves several assumptions regarding the IMFP. In theoretical IMFP reports, it has been pointed out that the predictive accuracy of IMFP for photoelectrons with kinetic energies below 10 eV is limited and strongly material-dependent [25,26]. A more rigorous analysis would require experimental verification of the IMFP for low-energy photoelectrons.

**IV. Conclusion**

We have developed the laser-PEEM system with ADECS which can measure both ferroelectric polarization contrast and electronic state distribution in ferroelectric devices, and measured HfO$_2$-based ferroelectric capacitors. Using an optimal $C_{st}$, we successfully obtained clear ferroelectric hysteresis loops for the HZO-based capacitors by ADECS. The *operando* laser-PEEM system was used to observe the fatigue and breakdown processes of $HfO_2$-based ferroelectric capacitors. We successfully visualized a spot responsible for the conduction path just after dielectric breakdown. Spectromicroscopy measurements indicate that defect states form near $E_F$, providing spectroscopic evidence that the conduction path is comprised of these defect states. The polarization contrast in the IZO/HZO/TiN capacitor was also successfully observed nondestructively. This study demonstrates that ferroelectric polarization contrast and conduction filaments can be visualized without removing the overlying electrodes by the *operando* laser-PEEM system.

**Acknowledgements**

We thank Y. Mizuno for her technical support of *operando* laser-PEEM measurements. We also thank K. Okazaki and T. Kondo for their fruitful discussion. This work was supported by Grants-in-Aid for Scientific Research (KAKENHI) (Grants Nos. 21H04549 and 23K13363) from the Japan Society for the Promotion of Science (JSPS). This work was partly supported by Foundation for Promotion of Material Science and Technology of Japan (MST Foundation) and by The Precise Measurement Technology Promotion Foundation. This work is dedicated to the memory of Professor Takanori Koshikawa, whose invaluable contributions and leadership have shaped our field. His commitment to advancing scientific knowledge and fostering international collaboration leaves a lasting legacy. We are grateful for his inspiration and guidance.

**Competing interests**

The authors declare no competing interests.

**Figures and captions**

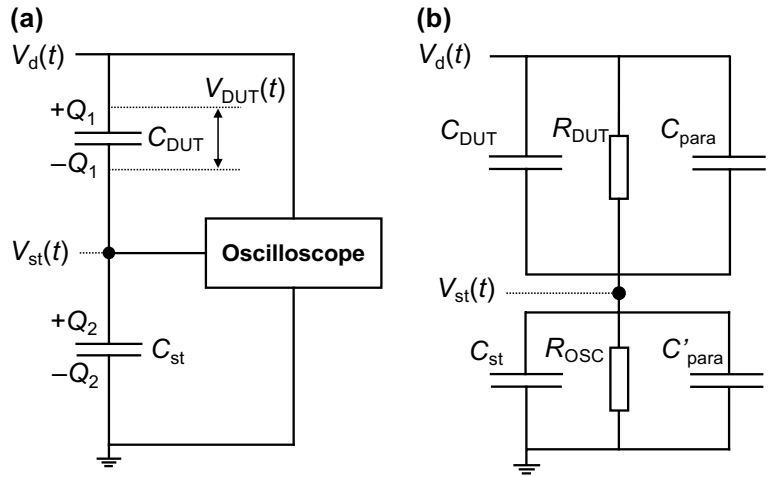

**Fig. 1** Schematic diagrams of the Sawyer-Tower (ST) circuit. **(a)** A simple ST circuit without parasitic components such as parasitic resistances, capacitances, and inductances. **(b)** A ST circuit including the resistive elements and the parasitic capacitances ($C_{para}$, $C'_{para}$) of the *operando* laser-PEEM system.

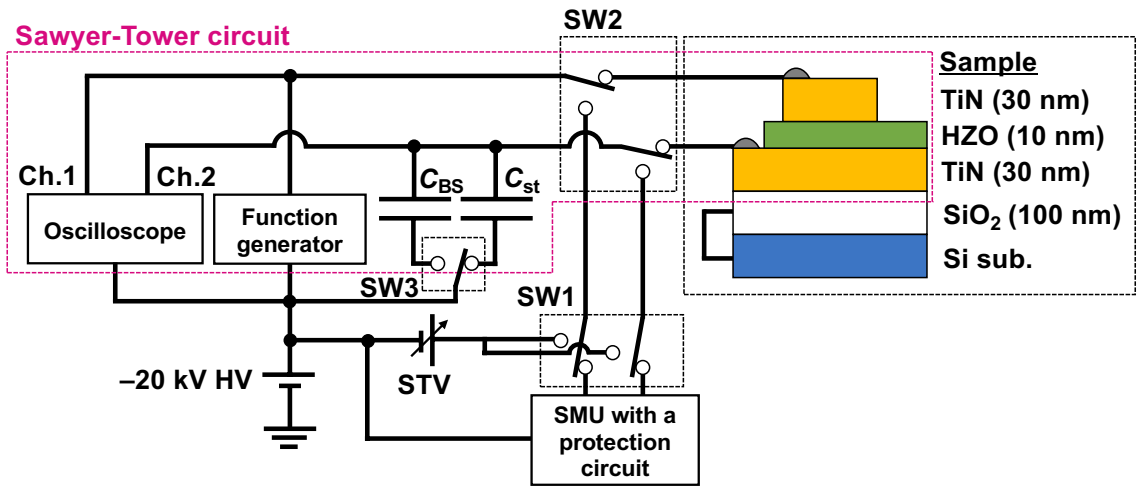

**Fig. 2** Schematic diagram of the electric system of the laser-PEEM system with ADECS including the sample capacitor.

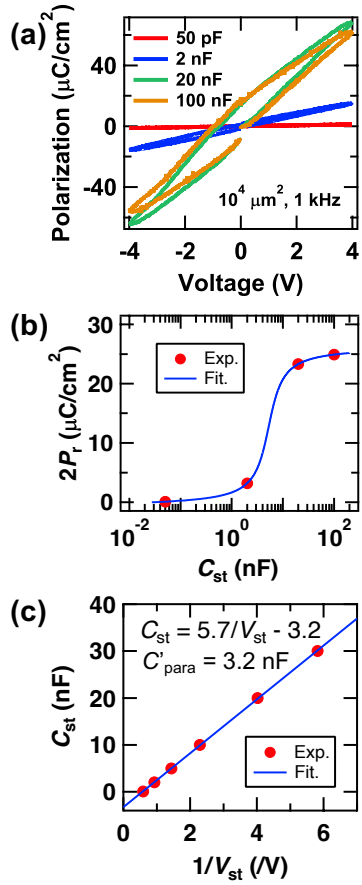

**Fig. 3** Estimation of the parasitic capacitance $C'_{para}$. **(a)** $C_{st}$ dependence of the observed $P$–$V$ hysteresis loop for DUT with an area of 100 μm × 100 μm. The measurements were performed by a single bipolar triangle wave at a frequency of 1 kHz. **(b)** Plot of $2P_r$ vs $C_{st}$ value. The curve fitting was performed by using a model $2P_r = A\arctan[B\log_{10}(C_{st}) - C] + D$, where A, B, C, and D indicate the fitting parameters. **(c)** Plot of $C_{st}$ vs the reciprocal voltage $1/V_{st}$ measured across $C_{st}$. Vst values were extracted as the voltage applied across $C_{st}$ when the drive voltage was 3 V.

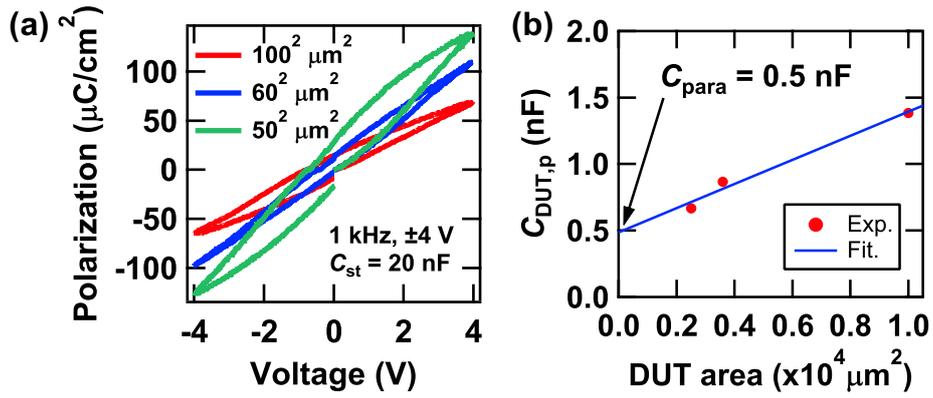

**Fig. 4** Estimation of the parasitic capacitance $C_{\text{para}}$. **(a)** DUT area dependence of the $P$–$V$ hysteresis loop. The measurements were performed by a single bipolar triangle wave at a frequency of 1 kHz. A condenser of 20 nF was used as the standard capacitor $C_{\text{st}}$. **(b)** Plot of $C_{\text{DUT,p}}$ vs DUT area.

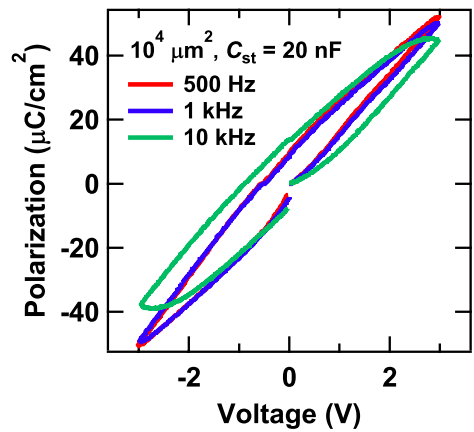

**Fig. 5** $V_{\text{d}}$-frequency dependence of the $P$–$V$ hysteresis loop taken for the DUT area of 100 μm ×100 μm and $C_{\text{st}}$ of 20 nF.

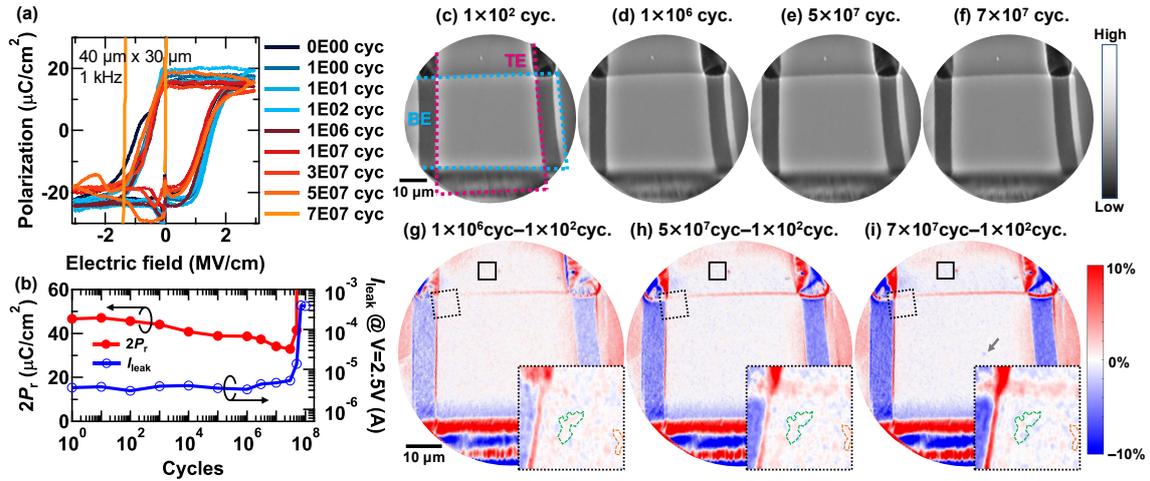

**Fig. 6** Field-cycling characteristics of the 40 μm × 30 μm capacitor with 10-nm thick HZO observed with –20-kV high voltage between the sample and the objective lens. **(a)** Evolution of $2P_r$ (red) and $I_{leak}$ (blue) with field cycling. $I_{leak}$ was measured by SMU implemented in the *operando* laser-PEEM system. **(b)** *P–V* hysteresis loops taken by PUND method for each number of field-cycling stress application. Noise was removed by a smoothing procedure. **(c)–(f)** PEEM image taken after $1\times10^2$, $1\times10^6$, $5\times10^7$, and $7\times10^7$ cycles, respectively. The field of view is approximately 50 μm. **(g)–(i)** Difference images obtained by subtracting the PEEM images shown in (c) from (d), (e), and (f), respectively. The differential intensities $I_{diff}$ for each pixel were calculated using the equation $I_{diff} = (I - I_{1E02cyc})/(I + I_{1E02cyc})$, where $I$ indicates the photoelectron intensities after $1\times10^6$, $5\times10^7$, or $7\times10^7$ cycles, and $I_{1E02cyc}$ indicates those after $1\times10^2$ cycles. Before subtracting the images, the intensities were normalized by the total intensity within the region shown by the black solid square. The insets show the enlarged views of the area enclosed by the black dotted squares. The green and orange dashed lines are discussed in the text.

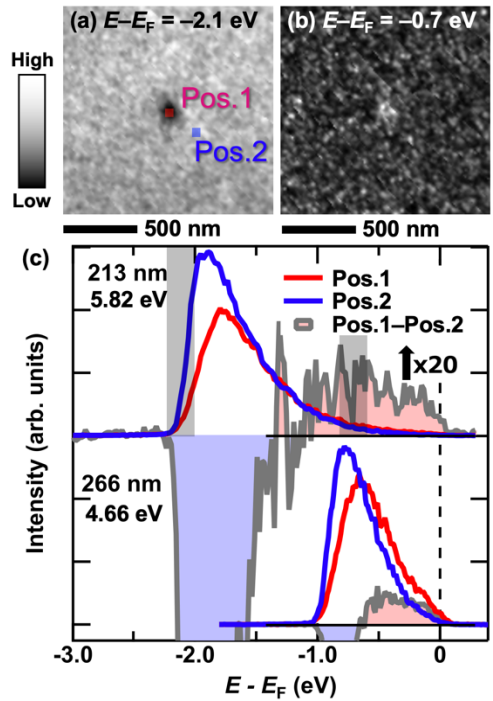

**Fig. 7** Photoelectron spectromicroscopy performed by using the energy slit and by sweeping STV. **(a)** Energy-filtered PEEM image at $E–E_F = –2.1$ eV. Red and blue squares indicate the intensity-integration region for extracting EDCs shown in (c). **(b)** Same as (a) but filtered at $E–E_F = –0.7$ eV. **(c)** EDCs extracted at Pos.1 and Pos.2 shown in (a), measured by 213-nm laser and 266-nm laser. Dark shaded regions correspond to the energies set in (a) and (b). The filled spectra show the differences between Pos.1 and Pos.2.

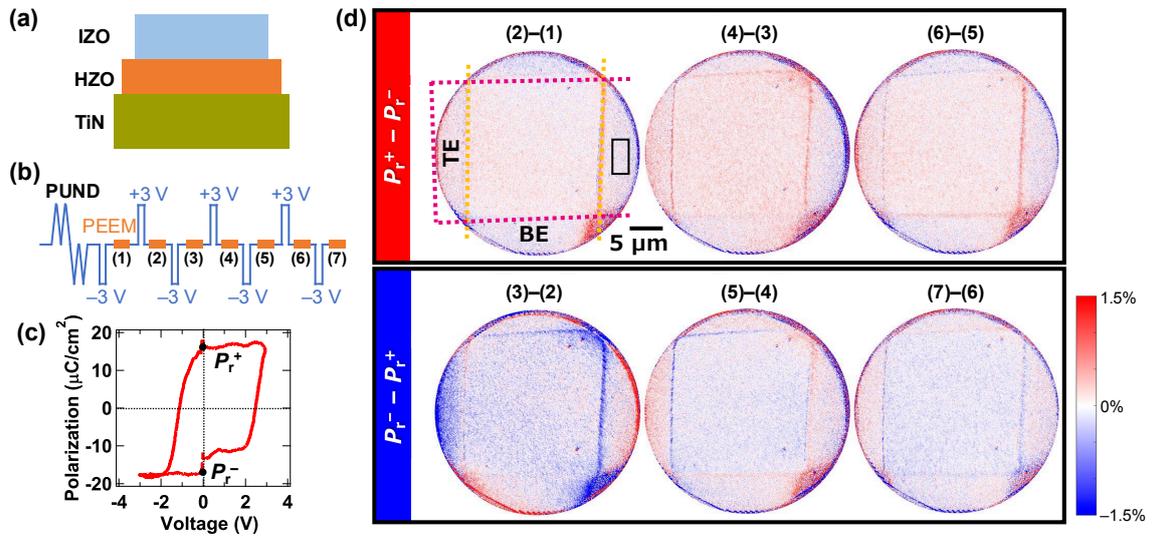

**Fig. 8** Polarization contrast in an IZO/HZO/TiN capacitor. **(a)** Cross-sectional schematic of the sample structure. **(b)** Measurement sequence. **(c)** $P$–$V$ hysteresis curve obtained by a PUND measurement. **(d)** Differential PEEM images to visualize the polarization contrast. The expression at the top of the images, e.g. (2)–(1), shows the difference between the PEEM image acquired at (2) defined in (b) and that acquired at (1), normalized by their sum.

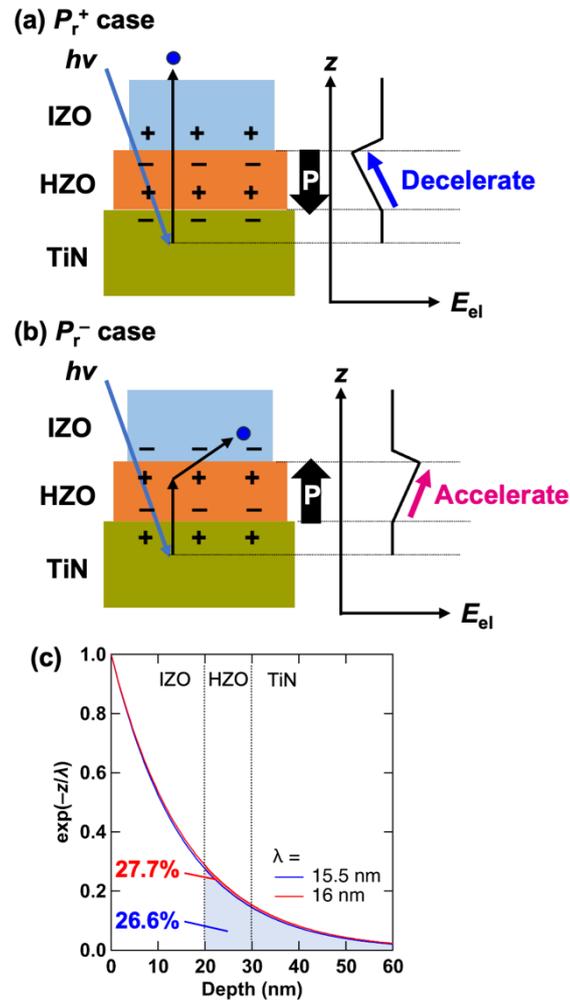

**Fig. 9** Possible contrast mechanism for the polarization contrast observed by PEEM. **(a)** Cross-sectional schematic of the capacitor including charge distribution in the $P_r^+$ state (left) and the energy evolution of photoelectrons excited in BE to reach the TE surface. **(b)** Same as (a) but for the $P_r^-$ state. **(c)** Estimation of the photoelectron intensity contribution from regions deeper than the IZO/HZO interface for IMFP of 16 nm and 15.5 nm. This analysis assumes that the IMFPs in IZO, HZO, and TiN are equivalent.

# Supplementary Materials for
# Breakdown and polarization contrasts in ferroelectric devices observed by *operando* laser-based photoemission electron microscopy with the AC/DC electrical characterization system

Hirokazu Fujiwara[1,2,†], Yuki Itoya[3], Masaharu Kobayashi[4], Cédric Bareille[5], and Toshiyuki Taniuchi[1,2]

**Affiliations:**

[1]*Department of Advanced Materials Science, Graduate School of Frontier Sciences, The University of Tokyo, Chiba 277-8561, Japan*

[2]*Material Innovation Research Center (MIRC), The University of Tokyo, Chiba 277-8561, Japan*

[3]*Institute of Industrial Science, The University of Tokyo, Tokyo 153-8505, Japan*

[4]*System Design Research Center (d.lab), School of Engineering, The University of Tokyo, Tokyo 153-8505, Japan*

[5]*Institute for Solid State Physics, The University of Tokyo, Chiba 277-8581, Japan*

† E-mail: hfujiwara@issp.u-tokyo.ac.jp

**1. V-t Waveforms Measured with the Sawyer-Tower Circuit**

When the capacitor exhibits high insulation resistance, the Sawyer–Tower (ST) circuit functions properly. As shown in the lower panel of Fig. S1(a) and in Fig. S1(b), the voltage across the standard capacitor, $V_{st}$, remains low, around 0.3 V, and the waveform shows distortion due to polarization switching. However, once dielectric breakdown occurs in the capacitor, the device under test (DUT) no longer functions as a capacitor. Consequently, the Sawyer–Tower circuit no longer represents a series connection of $C_{DUT}$ and $C_{st}$, but instead behaves as a series circuit composed of the resistance $R$ of DUT and $C_{st}$.

The resistance after breakdown is still higher (~10 kΩ) than that of the parasitic resistance (~1 kΩ). Moreover, the equivalent capacitance changes from $C_{DUT}C_{st} / (C_{DUT}+C_{st}) \sim C_{DUT}$ (since $C_{DUT} \ll C_{st}$) to approximately $C_{st}$. As a result, the time constant required to charge $C_{st}$, given by $\tau = RC_{st}$, is estimated to be approximately 0.2 ms (with $R = 10$ kΩ, corresponding to the post-breakdown resistance, and $C_{st} = 20$ nF).

Under these conditions, in the PUND (Positive-Up–Negative-Down) measurement, the first triangular voltage pulse (width: 0.25 ms) does not fully charge the capacitor $C_{st}$. Additional charging occurs during the second triangular pulse, leading to an increased voltage applied across $C_{st}$, as seen in the upper panel of Fig. S1(a). The difference in response between the first and second triangular pulses becomes significantly larger than that observed before breakdown. Consequently, the extracted polarization value is significantly overestimated, as illustrated in Fig. S1(d).

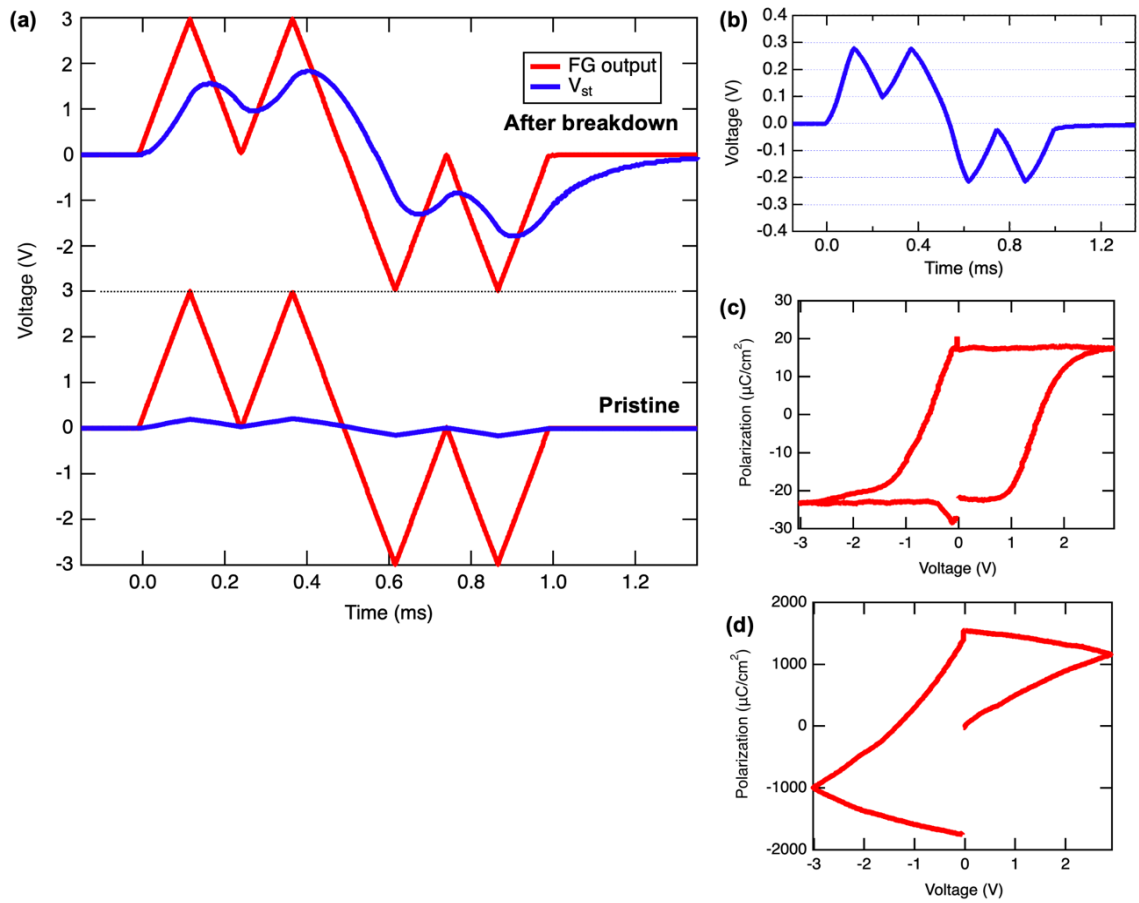

**Fig. S1 (a)** Output waveform from the function generator (FG) and the corresponding $V_{st}$ waveform measured for a capacitor after dielectric breakdown (upper panel) and for a pristine capacitor (lower panel). **(b)** Enlarged view of the $V_{st}$ waveform corresponding to the pristine capacitor shown in (a). **(c)** $P$–$V$ hysteresis loop obtained from the pristine capacitor. **(d)** Same as (c), but measured for the capacitor after dielectric breakdown.